\documentclass[prd,showpacs,twocolumn]{revtex4}

\newcommand{\beq}{\begin{equation}}
\newcommand{\eeq}{\end{equation}}
\newcommand{\beqar}{\begin{eqnarray}}
\newcommand{\eeqar}{\end{eqnarray}}

\newcommand{\prp}{\bot}
\newcommand{\del}{\partial}
\newcommand{\dell}[1]{\stackrel{\leftarrow}{\partial^{#1}}}
\newcommand{\dells}[1]{\stackrel{\leftarrow}{\partial_{#1}}}
\newcommand{\delr}[1]{\stackrel{\rightarrow}{\partial^{#1}}}
\newcommand{\dellr}[1]{\stackrel{\leftrightarrow}{\partial^{#1}}}
\newcommand{\dellrs}[1]{\stackrel{\leftrightarrow}{\partial_{#1}}}
\newcommand{\dellrss}[2]{\stackrel{\leftrightarrow}{\partial_{#1}^{#2}}}

\ifx\undefined\fiverm
     \font\fiverm=cmr5
\fi

\input{prepictex}
\input{pictex}
\input{postpictex}

\begin{document}
\author{Tarun Biswas}
\title{Another Alternative to the Higgs Mechanism}
\email{biswast@newpaltz.edu}
\affiliation{State University of New York at New Paltz, \\ New Paltz,  NY 12561, USA.}
\date{\today}
\begin{abstract}
The Higgs mechanism is designed to generate mass for massless particles. The mass
comes from the interaction of observed particles with an external field -- the Higgs field.
In the past, several alternatives to the Higgs mechanism for mass generation have been proposed 
to avoid the postulation of the Higgs field. This article proposes yet another one.
This alternative is distinctly different from the others because it
considers mass generation through {\em internal} interactions of a particle rather than 
interactions with external fields. This requires particles to have an internal structure beyond
intrinsic spin. A complete field theory of such composite particles is seen to be possible. 
Of course, if Higgs bosons are observed by experiment, 
there will be no need for any alternatives. On the other hand, if experiment fails to 
detect Higgs bosons, such alternate mechanisms for particle mass generation would be very useful.
\end{abstract}
\pacs{12.60.-i, 11.10.-z, 03.70.+k}
\maketitle
\section{Introduction}
The elegance of particle physics theory comes from the underlying symmetry groups. Larger
the number of particles that fall under one group, the better. Differences in masses of
particles can ruin such symmetry. Hence, the Higgs mechanism is found to be very useful.
It can generate mass dynamically, thus allowing particles to have zero rest mass in their
interaction-free equations of motion. However, the Higgs mechanism necessitates the existence of the
Higgs bosons which have not been experimentally detected so far. Hence, mass generation
without the postulation of the Higgs field is an interesting possibility. Several such mechanisms
have already been proposed\cite{hill,chivu,csaki,he}. The mechanism proposed here is different from the others.
It generates mass through {\em internal} interactions rather than external interactions with
other fields. This requires that internal structure be ascribed to traditional structureless
particles. Hence, a field theory of structured (or extended) objects needs to be developed. 
String theories deal
with such extended objects. However, strings having infinite degrees of freedom complicates
matters. The theory discussed here considers fields of extended objects with finite numbers of
internal degrees of freedom. To maintain relativistic covariance (both classical and quantum)
in the treatment of internal interactions of such extended objects one uses methods developed 
in relativistic hamiltonian constraint dynamics\cite{bisrohr, arens, droz, tod, kom, lusa, 
sazd, bidi}. These methods have been used successfully over many years in models for particle 
bound states\cite{bis0, crater, crater1}.

\section{Classical Extended Objects: Rest-Mass and Effective Mass Generation}

Classically, the mass-shell condition for a free particle is as follows.
\begin{equation}
	p^{2}+m^{2}=0,
\end{equation}
where $p$ is the four-momentum of the particle and $m$ its rest-mass. This is equivalent to stating
\begin{equation}
	p^{\|}=m, \label{eqfreemsh}
\end{equation}
where, for any four-vector $v$, the component parallel to the momentum (zeroth
component in the center-of-mass (CM) frame) is given by
\begin{equation}
	v^{\|}=-v\cdot\hat{p},\;\;\; \hat{p}=p/\sqrt{-p^{2}}
\end{equation}
So, $p^{\|}$ is the center-of-mass energy as well as the rest-mass of the particle. Hence, 
one may make the following rather trivial observation.
\begin{quotation}
	{\em Rest-mass is the energy in the CM frame.}
\end{quotation}

However, the same statement is not that trivial when applied to composite objects with 
interacting components.
Consider a point particle with another point object attached to it by some confining force.
We shall call the point particle the {\em vertex} (or the {\em bare particle}) 
and the attached point object the {\em satellite}.
Let the position and momentum four-vectors for the vertex be $q_{0}$ and $p_{0}$ respectively and
those for the satellite be $q_{1}$ and $p_{1}$ respectively. As we are interested in
a theory of massless bare particles, we need to focus on the rest-mass of the vertex. The
nature of this rest-mass is complicated by the fact that the vertex is bound to the satellite by 
a confining force. The rest-mass of a free particle is its energy at rest. But the vertex, by
itself, is never free. Hence, for the purpose of interacting objects like the vertex, 
the definition of rest-mass needs to be generalized to the above statement which happens to 
be trivial for free particles -- rest-mass is the energy in the CM frame.

For the vertex, its own CM frame is not inertial. So we use the CM frame
of the whole composite of vertex and satellite. The total momentum of the composite is
\begin{equation}
	P = p_{0}+p_{1}. \label{eqtotmom}
\end{equation}
So the energy of the vertex in the CM frame is denoted by the component of
$p_{0}$ parallel to the total momentum $P$:
\begin{equation}
	p_{0}^{\|}=-p_{0}\cdot\hat{P},\;\;\; \hat{P}=P/\sqrt{-P^{2}}. \label{eqpardef}
\end{equation}
Hence, the rest-mass for the vertex is as follows.
\begin{equation}
	p_{0}^{\|}=m.
\end{equation}
This is the equivalent of the mass-shell condition for free particles (equation~\ref{eqfreemsh}).

Now, for the sake of symmetry, if the rest-mass of the vertex (or the bare particle) 
were to be zero, it would have the following mass-shell condition.
\begin{equation}
	p_{0}^{\|}=0. \label{eq0rm}
\end{equation}
This can be rewritten (using equations~\ref{eqtotmom} and ~\ref{eqpardef}) as
\begin{equation}
	P^{2}+(p_{1}^{\|})^{2}=0,
\end{equation}
where $p_{1}^{\|}$ is the component of $p_{1}$ along $P$. Naturally, if $M$ were the 
{\em effective mass} of the whole composite, then
\begin{equation}
	P^{2}+M^{2}=0. \label{eqcmpmsh}
\end{equation}
Hence, we identify the dynamically generated effective mass as
\begin{equation}
	M\equiv p_{1}^{\|}. \label{eqcmpmass}
\end{equation}
 Clearly, this is generated by internal dynamics and can be non-zero while the rest-mass of
 the vertex is zero. $M$ can be seen to be a constant of motion that depends on initial conditions
 and the nature of the interaction between vertex and satellite.
To see this, one needs to find the relationship of $p_{1}^{\|}$ to the interaction of the vertex and
satellite.

Equation~\ref{eq0rm} is the mass-shell condition for the zero rest-mass vertex. Equivalently
(equation~\ref{eqcmpmsh}), it is also the effective mass-shell condition for the composite.
But the composite is made up of two point objects and hence, it must have two independent
mass-shell conditions as required by relativistic hamiltonian constraint dynamics\cite{bisrohr,
arens, droz, tod, kom, lusa, sazd, bidi}.
The second mass-shell condition is imposed on $p_{1}$, the momentum of the satellite as follows.
\begin{equation}
	p_{1}^{2}+m_{1}^{2}=0. \label{eqsmsh}
\end{equation}
If $m_{1}$ were just a constant, it would be the mass of the satellite and require
that the satellite be a free particle. But the satellite is not a free particle. So, $m_{1}$
cannot be a constant. Requiring that $m_{1}$ be a function of the position of the satellite 
relative to the vertex, effectively introduces the interaction between the vertex and the
satellite. The relative coordinate between the vertex and the satellite is
\begin{equation}
	\xi = q_{1}-q_{0}.
\end{equation}
So,
\begin{equation}
	m_{1}\equiv m_{1}(\xi).
\end{equation}
This also shows that
a confined particle (the satellite) can be viewed as something with a variable mass -- a mass
that increases with distance from the vertex.

Classical hamiltonian constraint dynamics requires that the two mass-shell conditions given by
equations~\ref{eqcmpmsh} and~\ref{eqsmsh} satisfy further conditions for consistency.
These conditions can be specified many different ways. However, for a smooth transition from
classical to quantum, the following Poisson bracket condition is preferred\cite{bisrohr,
tod, kom}.
\begin{equation}
	\{D_{0},D_{1}\}=0, \label{eqpbcom}
\end{equation}
where,
\begin{equation}
	D_{0}=P^{2}+M^{2},\;\;\;D_{1}=p_{1}^{2}+m_{1}^{2}. \label{eqd0d1}
\end{equation}
It is to be noted that although $D_{0}$ and $D_{1}$ individually vanish due to the mass-shell
conditions, their Poisson bracket can still be non-zero. Hence, equation~\ref{eqpbcom} is
a condition independent of the mass-shell conditions. It can be seen that this new 
condition is satisfied if the function $m_{1}(\xi)$ is restricted to be a function
of $\xi^{\prp}$ alone\cite{bisrohr, bis1} where, for an arbitrary four-vector $v$, we define
\begin{equation}
	v^{\prp}\equiv v\cdot P^{\prp},\;\;\; P^{\prp} \equiv \eta+\hat{P}\hat{P}.
\end{equation}
Here $\eta$ is the Minkowski metric and $P^{\prp}$ is the projection operator that
projects orthogonal to $P$. Hence,
\begin{equation}
	m_{1}\equiv m_{1}(\xi^{\prp}). \label{eqm1cnd}
\end{equation}
This states that $m_{1}$ must be a function of only the spatial components of $\xi$ in the
CM frame.

It can now be noticed that $\sqrt{-P^{2}}$ is the energy in the CM frame and that
\begin{equation}
	\{M,\sqrt{-P^{2}}\}=\{p_{1}^{\|},\sqrt{-P^{2}}\}=0.
\end{equation}
Hence, $M$ is a conserved quantity and it can be treated as the dynamically generated effective mass
of the composite as indicated by equation~\ref{eqcmpmsh}. As this effective mass is dynamically
generated, it will depend on initial conditions. Classically, $M$ can
acquire a continuum of values depending on initial conditions. However, in a quantized model
it can be restricted to certain discrete values. The following sections deal with the quantization
of this composite particle model.

\section{First Quantization of Extended Objects (Composite Particles)}
To first quantize the composite object described above, a convenient set of phase space 
variables needs to be identified. The most obvious set is the following.
\begin{equation}
	S_{p0}= \{p_{0},p_{1},q_{0},q_{1}\}.
\end{equation}
Quantization amounts to converting the Poisson bracket relations of this space to commutator
bracket relations. This gives the following non-zero commutator brackets.
\begin{equation}
	[q_{0},p_{0}]=i\eta,\;[q_{1},p_{1}]=i\eta.
\end{equation}
All other commutators are zero. A canonical transformation of $S_{p0}$ to accommodate the
translation invariant $\xi$ is useful. This produces the following phase space variables.
\begin{equation}
	S_{p1}= \{P,\pi,Q,\xi\},
\end{equation}
where
\begin{equation}
	P=p_{0}+p_{1},\;\pi=p_{1},\;Q=q_{0},\;\xi=q_{1}-q_{0}.
\end{equation}
The non-zero commutators of $S_{p1}$ are as follows.
\begin{equation}
	[Q,P]=i\eta,\;[\xi,\pi]=i\eta.
\end{equation}
A less trivial transformation is to the following set.
\begin{equation}
	S_{p}= \{P,\pi^{\|},\pi^{\prp},x,\xi^{\|},\xi^{\prp}\},
\end{equation}
where the components of $\pi$ and $\xi$ parallel and orthogonal to $P$ (CM components)
are used. However, in this set $Q$ cannot be used any more. This is because the CM components of
$\pi$ and $\xi$ depend on $P$ and hence their commutators with $Q$ do not vanish.
However, it can be proved\cite{bis2} that there exists an $x$ such that the only non-zero
commutators of $S_{p}$ are the following.
\begin{equation}
	[x,P]=i\eta,\;[\xi^{\|},\pi^{\|}]=-i,\;[\xi^{\prp},\pi^{\prp}]=iP^{\prp}. \label{eqcomr}
\end{equation}
As long as the existence of $x$ is proven, its explicit dependence on the variables 
of $S_{p1}$ is not necessary for the discussion of a quantum theory. Due to the
commutation relations, $x$ behaves as the position of the composite particle.

The commutation relations of $S_{p}$ provide the following maximal set of mutually
commuting variables.
\beq
S_{L}= \{x,\xi^{\|},\xi^{\bot}\}. \label{eq8}
\eeq
Hence, the first quantized wavefunction of the system can be
written as a function on $S_{L}$.
\beq
\psi = \psi(x,\xi^{\|},\xi^{\bot}). \label{eq9}
\eeq
The commutation conditions of equation~\ref{eqcomr} lead to the following
differential operator representation of the momenta.
\beq
i(-\pi^{\|},\pi^{\bot})=\del_{s\alpha}\equiv
\left(\frac{\del}{\del\xi^{\|}},\nabla\right), \label{eq10}
\eeq
and
\beq
iP_{\alpha}=\del_{\alpha}\equiv\frac{\del}{\del x^{\alpha}}, \label{eq11}
\eeq
where $\alpha$ is the four-vector index and the subscript $s$ identifies the derivatives
with respect to the satellite relative coordinates $\xi$. The
four-vector component notation $(\cdots,\cdots)$ gives the zeroth component
as the first argument and the three-vector components as the second argument.
The three-vector operator $\nabla$ is defined to be the gradient in the 
three-vector space of $\xi^{\bot}$ which represents the spatial components
of $\xi$ in the CM frame.
\beq
\nabla\equiv \left(\frac{\del}{\del\xi_{1}},\frac{\del}{\del\xi_{2}},
\frac{\del}{\del\xi_{3}}\right). \label{eq12}
\eeq
The wavefunction $\psi$ must satisfy one equation of motion for the
satellite and one for the vertex. These equations come from the classical
mass-shell conditions of equations~\ref{eqcmpmsh} and~\ref{eqsmsh}. Using the
first-quantized forms of $D_{0}$ and $D_{1}$ (equation~\ref{eqd0d1}), these
mass-shell conditions give the following quantum equations of motion.
\beqar
D_{0}\psi\equiv \left(\del_{\mu}\del^{\mu}-M^{2}\right)\psi & = & 0. \label{eq13}	\\
D_{1}\psi\equiv (\del_{s\alpha}\del_{s}^{\alpha}-m_{1}^{2})\psi & = & 0, \label{eq13a}
\eeqar
The operator form of $M$ comes from the quantum version of equation~\ref{eqcmpmass}.
\beq
M=\pi^{\|}=i\frac{\del}{\del\xi^{\|}}. \label{eq15}
\eeq

The consistency condition of equation~\ref{eqpbcom} translates to the following
quantum form.
\beq
[D_{0},D_{1}]\psi=0. \label{eq16}
\eeq
As expected, this produces the same condition on $m_{1}$ as seen in the classical case
(equation~\ref{eqm1cnd}). If the functional form for $m_{1}$ is chosen to produce a
confining effect on the satellite, the solutions of equations~\ref{eq13} and~\ref{eq13a}
will produce a discrete spectrum for the eigenvalues of $\pi^{\|}$ and hence, $M$.
This spectrum of values of $M$ represent the possible values of effective mass for the composite
particle. In a field theory, each eigenvalue of $M$ will represent a different particle.

\section{Second Quantization of Composite Particles}
For a second quantized theory of composite particles, it is important to notice that the vertex
and the satellite are never individually free. This makes it possible to second quantize the
whole composite without second quantizing the vertex or the satellite individually. When the
whole composite is second quantized, the internal dynamics of vertex and satellite can be
completely represented through quantum numbers for internal energy (given by $M$) and 
angular momentum. These quantum numbers can be visualized as extensions of the set of intrinsic 
particle quantum numbers like spin.

The recipe for second quantization of composite particles is a generalization of the usual canonical
second quantization procedure. We start by defining the
following {\em universal current} as a generalization of conserved currents
in theories of structureless particles.
\beq
j^{\mu\alpha}\equiv (1/4)\psi^{\dagger}\dellr{\mu}\dellrss{s}{\alpha}\psi, \label{eq18}
\eeq
where $\psi^{\dagger}$ is the adjoint of $\psi$ and $\dellr{\mu}$ is defined by the following.
\beq
\psi^{\dagger}\dellr{\mu}\psi\equiv \psi^{\dagger}(\delr{\mu}-\dell{\mu})\psi \equiv
\psi^{\dagger}(\del^{\mu}\psi)-(\del^{\mu}\psi^{\dagger})\psi. \label{eq19}
\eeq
$\dellrss{s}{\alpha}$ is defined similarly using the satellite relative coordinates.
This yields the following conserved currents.
\beq
j^{\mu}\equiv \int j^{\mu\alpha}d^{3}\xi_{\alpha}.\;\;
j_{s}^{\alpha}\equiv \int j^{\mu\alpha}d^{3}x_{\mu},
\label{eq21}
\eeq
where terms like $d^{3}x_{\mu}$ represent four-vector hypersurface elements in
$x_{\mu}$ space. The integrations are done over arbitrary infinite spacelike 
hypersurfaces. It is straightforward to prove the following conservation equations using 
the equations of motion\cite{bis2}.
\beq
\del_{s\alpha}j_{s}^{\alpha}=0, \label{eq22}
\eeq
and
\beq
\del_{\mu}j^{\mu}=0. \label{eq23}
\eeq
Both conserved currents lead to the same conserved charge. It is given by
\beq
{\cal Q}\equiv \int j^{\mu\alpha}d^{3}\xi_{\alpha}d^{3}x_{\mu}.
\label{eq24}
\eeq
Due to the conservation equations~\ref{eq22} and~\ref{eq23} it can be seen that the
two integrations over space-like hypersurfaces are independent of the choice of
any specific hypersurface. Hence, for convenience, we choose the $\xi_{\alpha}$
hypersurfaces to be orthogonal to the total momentum $P$. This makes sure it is
purely spatial in the CM frame. So, we replace $d^{3}\xi_{\alpha}$ by
$d^{3}\xi^{\bot}$ and $\dellrss{s}{\alpha}$ by $\dellrss{s}{\|}$ where
\beq
\dellrss{s}{\|}\equiv -\hat{P}_{\alpha}\dellrss{s}{\alpha}. \label{eq25}
\eeq
For the $d^{3}x_{\mu}$ integration we choose the purely spatial components ${\bf x}$
in the laboratory frame. Hence, $\dellr{\mu}$ can be replaced by $\dellr{0}$. This
gives the conserved charge to be
\beq
{\cal Q}\equiv (1/4)\int\psi^{\dagger}\dellrs{0}\dellrss{s}{\|}
d^{3}\xi^{\bot}d^{3}{\bf x}. \label{eq26}
\eeq
This conserved charge suggests the following natural norm for the Hilbert space of
$\psi$.
\beq
(\psi,\psi)\equiv 1/4\int\psi^{\dagger}\dellrs{0}\dellrss{s}{\|}\psi
d^{3}\xi^{\bot}d^{3}{\bf x}. \label{eq27}
\eeq
This leads to the following definition of the inner product.
\beq
(\phi,\psi)\equiv 1/4\int\phi^{\dagger}\dellrs{0}\dellrss{s}{\|}\psi
d^{3}\xi^{\bot}d^{3}{\bf x}. \label{eq28}
\eeq
The above inner product definition allows us to identify the following
orthonormal basis for the set of solutions of the equations of motion.
\begin{equation}
\psi_{{\bf k}E} \equiv [k^{0}(2\pi)^{3}]^{-1/2}\Psi(E,\xi^{\bot})
\exp[-iE\xi^{\|}]\exp(ik\cdot x), \label{eq29}
\end{equation}
where $k$ is the four-vector eigenvalue of the total momentum $P$, ${\bf k}$
is its three-vector part and $k^{0}$ is its zeroth component. $E$ is the
CM energy of the satellite and hence, an eigenvalue of $\pi^{\|}$ or $M$.
Note that $E$, can be negative. This requires the usual
explanation of an antiparticle being a particle going backward in time. So the
physically measurable mass is still positive.
For $\psi_{{\bf k}E}$ to be a solution of the equations of motion
in the satellite sector, $\Psi(E,\xi^{\bot})$
must satisfy the following eigenvalue equation.
\beq
H\Psi(E,\xi^{\bot})=
E\Psi(E,\xi^{\bot}), \label{eq30}
\eeq
where
\beq
H\equiv \sqrt{(\pi^{\bot})^{2}+m_{1}^{2}}, \label{eq31}
\eeq
It is to be noted that $\Psi(E,\xi^{\bot})$ also
depends on angular momentum quantum numbers due rotational symmetry.
The labels for these quantum numbers are suppressed for brevity of notation. Also,
the spectrum for $E$ is expected to be discrete as $m_{1}$ produces a confining effect.
For $\psi_{{\bf k}E}$ to be a solution of the whole particle equation
of motion, the following must be satisfied.
\beq
k^{0}=\sqrt{{\bf k}^{2}+E^{2}}. \label{eq32}
\eeq
$\Psi$ may be normalized in the usual fashion.
\begin{equation}
	\int\Psi^{\dagger}(E',\xi^{\bot})\Psi(E,\xi^{\bot})d^{3}\xi^{\bot}= \delta_{E'E},
\label{eq33}
\end{equation}
where $\delta_{E'E}$ is the Kronecker delta and, once again, the angular momentum 
labels are suppressed and understood to be included in the corresponding energy
label. Using these
conditions, it can be verified that the $\psi_{{\bf k}E}$ are truly
orthonormal.
\beq
(\psi_{{\bf k}'E'},\psi_{{\bf k}E})=
\delta_{E'E}\delta({\bf k}'-{\bf k}), \label{eq34}
\eeq
where $\delta({\bf k}'-{\bf k})$ is the Dirac delta.

Now we are ready for second quantization. The standard prescription for
canonical quantization will be used. However, it is critical to note that
the satellite and the vertex are not second quantized individually. It is the whole
particle wavefunction $\psi$ that is second quantized. The energy and
angular momentum of the satellite are treated as extra degrees of freedom
(quantum numbers) of the whole particle wavefunction.

First, a Lagrangian for the particle field is defined as follows.
\beq
{\cal L}=-1/2\int\psi^{\dagger}\dellrss{s}{\|}[\dells{\mu}\delr{\mu}+M^{2}]\psi
d^{3}\xi^{\bot}, \label{eq35}
\eeq
The momentum conjugate to $\psi$ would then be
\beq
\phi\equiv\frac{\del{\cal L}}{\del(\del_{0}\psi)}=\del_{s}^{\|}\del^{0}\psi^{\dagger}, \label{eq36}
\eeq
Then the second quantization condition can be written symbolically as the following 
equal-time commutator\footnote{The direct quantization of classical composite particles
allows only bosonic particles. However, the model can be generalized to fermionic composites
as well\cite{bis2, bis3, bis4}}.
\beq
[\psi,\phi]=i\delta, \label{eq38}
\eeq
where the $\delta$ is a delta function over all degrees of freedom.

Now, $\psi$ can be expanded in terms of the basis set of equation~\ref{eq29} as follows.
\beqar
\psi & = & \int d^{3}{\bf k}\sum_{E}[2k^{0}(2\pi)^{3}]^{-1/2}
\Psi(E,\xi^{\bot})\exp[-iE\xi^{\|}]\cdot \nonumber \\
& & \cdot[b({\bf k},E)\exp(ik\cdot x)+d^{\dagger}({\bf k},E)\exp(-ik\cdot x)].
\label{eq39}
\eeqar
As in usual field theories, the $b$ and $d^{\dagger}$ coefficients are used to separate particle
and antiparticle states. $d^{\dagger}$ represents the hermitian adjoint of $d$ in a field operator
sense. Then, the quantization condition of equation~\ref{eq38}
reduces to the following (as before, the energy labels are understood to include
angular momentum labels).
\beqar
[b({\bf k},E),b^{\dagger}({\bf k}',E')] & = & [d({\bf k},E),d^{\dagger}({\bf k}',E')]= \nonumber \\
& = & \delta^{3}({\bf k}-{\bf k}')\delta_{E'E},
\label{eq40}
\eeqar
and all other commutators of $b$, $b^{\dagger}$, $d$ and $d^{\dagger}$ vanish. This allows the
building of the usual Fock space with $b^{\dagger}$ being the particle creation operator,
$b$ the particle annihilation operator, $d^{\dagger}$ the antiparticle creation operator
and $d$ the antiparticle annihilation operator. The necessary
vacuum state can be shown to be stable\cite{bis2}.

This is a field theory of a composite particle with a bosonic vertex and a bosonic satellite.
It is possible to generalize this to composites with vertex and satellite each being
either bosonic or fermionic\cite{bis2, bis3, bis4}. It is also possible to have multiple satellites.

\section{Conclusion}
An unusual mechanism for mass generation is discussed here. It requires the postulation of an
internal structure for particles. A satellite permanently attached to the bare particle is seen
to generate mass dynamically. Hence, this satellite may be considered to be a first quantized
equivalent of the Higgs boson. However, the satellite, being attached to the bare particle by
confining forces, is not expected to be detected independently.

\end{document}